\documentclass[12pt]{article}
\usepackage[utf8]{inputenc}
\usepackage{amsmath,amssymb,amsfonts,amsthm}
\usepackage{graphicx}
\usepackage{graphicx,psfrag,subfigure,color}
\usepackage{bbm}
\usepackage[caption=false]{subfig}
\usepackage{epstopdf}
\usepackage{color}
\usepackage{amsmath,amsthm,amsfonts,amssymb,bbm}
\usepackage{graphicx,psfrag,subfigure,color}
\usepackage{cite}

\numberwithin{equation}{section}

\newcommand{\e}{\epsilon}

\begin{document}
\title{The one-dimensional KPZ equation and its universality class }

\author{Jeremy Quastel\footnote{Department of Mathematics, University of Toronto, 40 St. George Street, Toronto, Ontario, Canada M5S 2E4. Email:  \texttt{quastel@math.toronto.edu}} 
\,\,and Herbert Spohn\footnote{Zentrum Mathematik and Physik Department,
Technische Universit\"at M\"unchen,
Boltzmannstra{\ss}e 3, 85747 Garching, Germany. Email: \texttt{spohn@ma.tum.de}}}

\date{March 2, 2015}

\maketitle
\vspace{2cm}
\begin{abstract} Our understanding of the one-dimensional KPZ equation, \textit{alias} noisy Burgers equation, has advanced substantially over the past five years. We provide a non-technical review,
where we limit ourselves to the stochastic PDE and lattice type models approximating it. 
\end{abstract}

\newpage
\section{Introduction}
\label{sec1}
 
Recently there has been spectacular progress on exact solutions for strongly interacting stochastic particle systems in one spatial dimension. Very roughly, these are two-dimensional field theories with some similarity to two-dimensional systems of equilibrium statistical mechanics and $1+1$ dimensional quantum field theories, including quantum spin chains. In these stochastic models the static correlations decay exponentially, but, because of conservation laws, the dynamic behavior shows non-trivial scaling properties.
 The physical interest in such models results from being representatives of huge universality classes. This is in direct analogy to the two-dimensional Ising model. Its exact solution is restricted to nearest neighbor couplings and zero external magnetic field. But the thereby predicted critical behavior is representative for a much larger class of models, including real two-dimensional ferromagnets to which, at first sight, the Ising model would be a very crude approximation.
 
 Our review attempts to answer ``What is currently known about the one-dimensional Kadar-Parisi-Zhang equation?''.
 This will cover mathematically rigorous results, but also important conjectures and intriguing replica computations.
 In addition, we discuss discrete models which in a particular scaling limit converge to the KPZ equation. They constitute 
 the universality class of the KPZ equation alluded to in the title. Such models have been, and still are, an important tool to arrive at information on the KPZ equation itself. Our program has obvious disadvantages. It mostly ignores in which temporal order discoveries were made. It also misses largely the amazing progress on other models in the KPZ universality class.
 Fortunately there are excellent reviews available \cite{Jo05,Sp06,FeSp11,CorwinRev,QRev,BoGo12,BoPe14,CorwinICM}, which cover further aspects of the story. 
 \medskip\\
\textbf{1.1 Interface motion}. In our article we focus on the KPZ universality class for growing fronts and interfaces in two-dimensional bulk systems.
With the hindsight of theoretical activities over many years, let us first recall the physical set-up under consideration.
One considers a thin film, for which two bulk phases can be realized such that they are in contact along an interfacial curve.
One of the phases is stable, while the other one is metastable with a very long life time. The dynamics of the bulk phases  has no conservation laws and is rapidly space-time mixing. Therefore, while  fluctuating, the bulk phases are not changing in time  
and the only observable motion is at the interface, where the transition from metastable to stable is fast,
while the reverse transition is very much suppressed. Thereby the stable phase is expanding into the metastable one,
accompanied by  characteristic interfacial fluctuations. If both phases would be   stable,
then there is no net motion, and the interface fluctuations are of size $t^{1/4}$ with Gaussian statistics on a sufficiently coarse scale. In the KPZ set-up on the other hand, in the frame where the bulk phases are at rest, the interface has a non-zero net velocity and fluctuations are of size $t^{1/3}$, a phenomenon known as kinetic roughening. There is no fine tuning of parameters. 
Any asymmetry in the growth process will lead to the $t^{1/3}$ scaling. Such motion comes with interfacial fluctuations which have non-Gaussian statistics, hence 
 the prime interest in exact solutions. 

The most prominent physical realization is a thin film of turbulent liquid crystal \cite{TS12,TSSS12}. 
In fact, the bulk phases are driven by a
time oscillating, spatially uniform electric field,  and thus not in thermal equilibrium. At a carefully chosen point in the phase diagram, 
the DSM2 phase (stable) is coexisting with the DSM1 phase (metastable). In the experiment, a point seed of the stable phase is planted in the bulk metastable phase and the growth of the DSM2 cluster is observed. In further series of experiments  also the case of a stable line seed is realized.
Another example is Eden cluster growth. Here one starts from a point seed (stable phase) and randomly adds extra
material to the surface of the already existing cluster. Hence the metastable phase is just ``all empty'' and the precise ballistic deposition mechanism is not specified by  the model. On a large scale, the cluster has an approximately circular shape. But there are substantial shape fluctuations, which have the same statistics as observed for the turbulent liquid crystal  \cite{AOF11,Tak12}.

\begin{figure}
\begin{center}
\psfrag{h}[l][b]{$h(j,t)$}
\psfrag{x}[c][b]{$j$}
\psfrag{stable}[l][b]{stable phase}
\psfrag{meta}[l][b]{metastable phase}
\includegraphics[height=5cm]{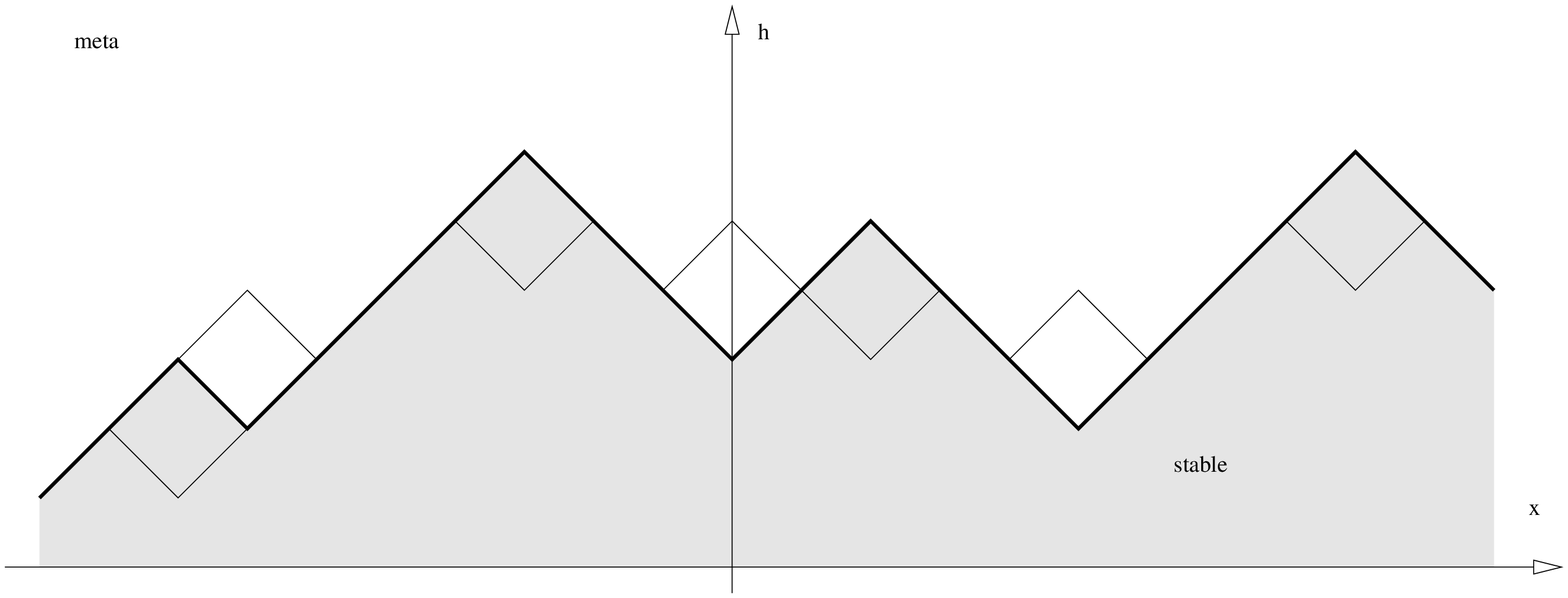}
\caption{The bold line is the current height. The transition from $\vee$ to $\wedge$ occurs with rate $q$, from $\wedge$ to $\vee$ with rate $p$, $q>p$.}
\label{fig1}
\end{center}
\end{figure}

Is there a model system which captures the essential features and is still amenable to a theoretical analysis?
The arguable simplest model is  the two-dimensional ferromagnetic Ising model with Glauber dynamics at zero temperature and a
suitably scaled positive magnetic field. The $+$ phase is stable and the $-$ phase is metastable. We assume that the $+$ phase is bordered from above by a lattice height function,
$h(j,t)$, which
is a broken line with slope $\pm 1$, see Fig. \ref{fig1}. The admissible growth sites are the local minima of $h$, at which 
independently a $-$ spin
flips to $+$ spin after an exponentially distributed waiting time, the reverse process occurring with a much smaller rate. In principle, clusters of the ``wrong'' sign could be spontaneously generated inside the bulk phases. But at low temperatures this is a very unlikely event and is completely suppressed at zero temperature. If instead of the height
one considers the local slope with value $\pm1$, then the interface dynamics amounts to a
random exchange of a neighboring $-+$ pair, the reverse process being less likely. This Markov jump process does not satisfy detailed balance, which is one reason for the unexpected dynamical behavior. The one-sided growth limit, $q =1$,
is known as TASEP, which has been a useful tool in obtaining exact solutions, see \cite{FeSp11}. 

Kardar, Parisi, and Zhang \cite{KPZ86} pose the same question, but proceed along a very different route, motivated by the Ginzburg-Landau theory of critical dynamics. Firstly, on a mesoscopic scale the interface is modeled as the graph of a height function $h(x,t)$ with time 
$t \geq 0$ and space $x \in \mathbb{R}$. In the case of contact between two stable phases one writes down the
Langevin equation
\begin{equation}\label{1.1}
\partial_t h = \nu \partial_x^2 h +\sqrt{D}\xi\,.
\end{equation}
Here $\nu$ is the surface tension and $\sqrt{D}\xi$ is a noise term of strength $\sqrt{D}$ describing the random back and forth between the two phases. $\xi(x,t)$ is modeled as space-time white noise, which 
is Gaussian with mean zero and covariance 
\begin{equation}\label{whitenoisecov}
\langle \xi(x,t)\xi(x',t') \rangle = \delta(x-x')\delta(t-t')\,.
\end{equation} 
On the basis of (\ref{1.1}) one finds that, if initially flat, the interface develops fluctuations of size $t^{1/4}$.

On the other hand, if the interface is stable-metastable, the net motion is modeled  by adding to (\ref{1.1}) a nonlinear drift, which is allowed to depend only on $\partial_x h$
so to maintain the invariance under the global shift of $h$ to $h +a$. For isotropic growth in the plane the nonlinearity can be easily guessed.
If locally there is a linear piece, $h(x,t) = a + \vartheta x$, then
$h(x, t + dt) - h(x,t) = v_\infty(1 + \vartheta^2)^{1/2}dt = F(\vartheta)dt$ with $v_\infty$ the asymptotic radial growth velocity. The precise form of the nonlinearity should not matter on large scales and, expanding at $\vartheta=0$,
one arrives at the one-dimensional KPZ equation
\begin{equation}\label{1.2}
\partial_t h =  \tfrac{1}{2}\lambda(\partial_x h)^2 + \nu\partial_x^2 h + \sqrt{D}\xi
\end{equation}
with $\lambda = F''(0)$.
In principle, we should keep the strength of nonlinearity, diffusion, and white noise  as free parameters, but by rescaling $x,t,h$ the conventional choice $\lambda = D=1$,  $\nu = \tfrac{1}{2}$, or any other, can always be achieved. Also the sign of the nonlinearity plays no role as can be
seen from the switch of $h$ to $-h$. \medskip\\
\textbf{1.2 Cole-Hopf solution}. The continuum description (\ref{1.2}) of the interface has its mathematical price. The solution to the linear equation
(\ref{1.1}) is locally a Brownian motion in $x$, i.e. it has everywhere cusp singularities as $\pm \sqrt{|x|}$. Hence  $\partial_xh$ of \eqref{1.2} is a distribution whose square does
not easily make sense. Fortunately there is a simple and instructive way out, which however is special for the quadratic nonlinearity. To keep things simple, set $\lambda=1 = D$, $\nu=1/2$ and introduce the Cole-Hopf transformation
\begin{equation}\label{1.3}
Z(x,t) = \mathrm{e}^{h(x,t)}\,.
\end{equation}
Then $Z$ is the solution of the stochastic heat equation,
\begin{equation}\label{1.4}
\partial_t Z(x,t) = \tfrac{1}{2} \partial_x^2Z(x,t) + \xi(x,t)Z(x,t)\,.
\end{equation}
The additive noise in (\ref{1.2}) has been turned into multiplicative noise. Using the Feynman-Kac formula, the ``solution'' to
(\ref{1.4}) with initial condition $Z(x,0)$ can
be written as
\begin{equation}\label{1.5}
Z(x,t) = \mathbb{E}_x\left[  \mathrm{e}^{\int_0^t ds~ \xi(b(s),s)} Z(b(0),0)\right]\,.
\end{equation}
Here $\mathbb{E}_x[\cdot]$ refers to the expectation w.r.t. an auxiliary standard Brownian motion  $b(s), 0 \leq s \leq t,$
with $b(t) = x$.
Despite $b(s)$ being continuous, as written, the integral in the exponential is ill-defined, since $\xi(x,t)$ is too rough. Still we expand the exponential and carry out the Brownian motion average with the result
\begin{eqnarray}\label{1.6}
&&\hspace{-30pt}Z(x,t) = \sum_{n = 0}^\infty   \int_{\mathbb{R}}dy\int_{0 < t_1< ...< t_n < t} dt_1 ... dt_n \nonumber\\
&&\hspace{15pt}\times \prod_{j =1}^n\Big( \int_{\mathbb{R}}dx_j \xi(x_j,t_j)\Big)
p_t(y,0; x_1,t_1, \ldots ,x_n,t_n;x,t) Z(y,0)\,.
\end{eqnarray}
Here $p_t$ is the time-ordered multi-time transition probability of Brownian motion. Eq. (\ref{1.6}) is a multiple stochastic integral, and we choose the forward (It\^o) discretization in every $t_j$. Then each
summand of (\ref{1.6}) is a well defined random variable, the sum converges in $L^2(\xi)$, and $Z(x,t) > 0$. We thus \textit{define} the (Cole-Hopf) solution to the KPZ equation as
\begin{equation}\label{1.7}
h(x,t) =  \log Z(x,t) \,.
\end{equation}
While the sum \eqref{1.6} seems to be useful only for studying the short time behavior of the solution, it does provide us with a well-defined starting point. 

The singular structure  of the KPZ equation comes from insisting on spatial white noise. If one would replace $\xi(x,t)$ by the smoothed version
$\xi_\varphi(x,t) = \int \varphi(x - x')\xi(x',t)dx'$ with $\varphi \geq 0$, even, smooth, of rapid decay at infinity, and normalized to $1$, then the solution to \eqref{1.2}
is well-defined and given by  \eqref{1.7} with the random partition function as defined in 
\eqref{1.5}. We now introduce an ultraviolet cut-off of size $\epsilon$ by choosing the $\delta$-function sequence  $\varphi_\epsilon(x) = \epsilon^{-1}\varphi(\epsilon^{-1}x)$ and  substituting the noise $\xi$ by its smoothed version $\xi_\epsilon = \xi_{\varphi_\epsilon}$. Such solution is denoted by $h_\epsilon(x,t)$. As established some time ago  \cite{BC,BeGi97}, in the limit $\epsilon \to 0$
one recovers the Cole-Hopf solution, provided one switches to a moving frame as $h_\epsilon(x,t) - v_\epsilon t$ with velocity $v_\epsilon  = \epsilon^{-1}\int \varphi(x)^2dx$ diverging in the limit. In field theory language this amounts to an infinite energy renormalization as 
familiar from $P(\phi)_2$ quantum field theory. 

In this context Hairer recently developed a solution theory, for which he was awarded the 2014 Fields Medal.
His theory works for the KPZ equation, as well as a large class of other singular stochastic partial differential equations,   and for general cutoffs \cite{Ha13,Ha13a}. For a specific cut-off, which does not pass nicely through the Cole-Hopf transformation, a somewhat more straightforward analysis can be found in \cite{Ku14}.  Of course,  the solution theory studies the small scale structure of solutions, and not the large scale, at which the universal behavior of interest is observed.
\medskip\\
\textbf{1.3 Directed polymers in a random medium}. Before diving into further details, let us recall an important observation made already in the 1986 paper.
To be specific, let us choose the initial condition $Z(x,0) = \delta(x)$. Then $b(s), 0 \leq s \leq t$ in \eqref{1.5} can be regarded as a directed polymer starting at time $0$ at $0$ and ending at time $t$ at $x$. It has been assigned  the random energy 
\begin{equation}\label{1.7aa}
\int_0^t ds\, \xi (b(s),s)
\end{equation}
as the sum over the potential $\xi$ along the path $b(s)$. With this perspective, $Z(x,t)$ is a partition function, the sum
being over all directed polymers. $Z(x,t)$ is random, because of the random potential $\xi$. Thus we really study a
disordered system from equilibrium statistical mechanics. The height of the interface, the object of physical interest,
is simply the (random) free energy. To leading order it is proportional to $t$, which corresponds to the non-random displacement of the height.
Thus physically our goal is to determine the fluctuations of the free energy. In the 
theory of spin glasses such fine details are rarely studied.\medskip\\
\textbf{1.4 Invariant measures and scaling}. If one introduces the slope $u = \partial_x h$, then $u$ satisfies the 
conservation law
\begin{equation}\label{1.7a}
\partial_t u+\partial_x\big(-\tfrac{1}{2}\lambda u^2  -\nu \partial_x u- \sqrt{D} \xi \big) = 0\,.
\end{equation} 
The inviscid Burgers equation corresponds to the case $\nu = 0$, $D = 0$. Eq. \eqref{1.7a} is hence referred to as noisy, or stochastic, Burgers equation. The linear equation, $\lambda = 0$, defines a Gaussian process. It has a 
one-parameter family of stationary measures, which are labeled by the mean $\langle u(x)\rangle = \vartheta$ and otherwise are Gaussian white noise with covariance $\langle u(x)u(x')\rangle - \vartheta^2 = \sigma^2\delta(x - x')$,
$\sigma = \sqrt{D/2\nu}$.   
Surprisingly enough, adding the nonlinearity does not modify the invariant measures. For this property to hold one only has to check that Gaussian white noise is invariant under the flow generated by solution to the deterministic 
evolution equation $\partial_t u= \tfrac{1}{2}\lambda\partial_x  u^2$. Formally its vector field is divergence free and the invariance follows from considering the time change of the action,
\begin{equation}\label{1.7b}
\frac{d}{dt} \int u ^2dx =  \lambda\int  u\partial_x u^2 dx=\tfrac13 \lambda\int \partial_x u^3 dx=  0.
\end{equation}

Our result can be translated to the KPZ equation \cite{FuQu14}. One chooses as initial conditions the two-sided Brownian motion $\sigma B(x) + \vartheta x$ with arbitrary slope $\vartheta$, \textit{i.e.} $B(0) = 0$ and $\{B(x), x \geq 0\}$, $\{B(x), x \leq 0\}$ are two independent standard Brownian motions. Then the solution $h(x,t)$ at fixed time $t$ is again two-sided Brownian motion with slope $\vartheta$ except for a random displacement, $h(x,t) - h(0,t) 
=  \sigma\widetilde{B}(x)+ \vartheta x$
in distribution with $\widetilde{B}(x)$ another two-sided Brownian motion. Note that $h(0,t)$ and $\widetilde{B}(x)$
are not independent.

With this observation one easily determines the scale relevant for fluctuations. We consider the solution 
$h(x,t)$ to \eqref{1.2} and define the rescaled fluctuation field
\begin{equation}\label{scal}
h_\e (x,t) = \e^b h(\e^{-1} x, \e^{-z} t )\,.
\end{equation}\label{1.7c}  
Using the scale invariance of white noise yields
\begin{equation}\label{scaling3}
\partial_t h_\e =\tfrac12\lambda \e^{2-z-b} (\partial_x h_\e)^2 +  \nu\e^{2-z} \partial_x^2 h_\e + \e^{b-\frac12 z+\frac12} \sqrt{D}\xi\,.
\end{equation}
To keep the stationary measure unchanged forces one to have the fluctuation exponent
$b=1/2$.
Thus to maintain the nonlinearity of order one requires
 \begin{equation}\label{1.7d}
 z=3/2\,.
 \end{equation} 
 Our rescaled equation takes the form 
\begin{equation}\label{1.7e}
\partial_t h = \tfrac12\lambda(\partial_x h)^2  +  \e^{1/2}\nu \partial_x^2 h  +  \e^{1/4} \sqrt{D} \xi\,.
\end{equation}
This scaling $h\sim \e^{-1/2}$, $t\sim \e^{-3/2}$, $x\sim \e^{-1}$ is more commonly thought of as 
$
h\sim t^{1/3}$, $x\sim t^{2/3}$.
This is the scale in which one observes the non-trivial fluctuation behavior throughout the KPZ universality class.  As to be noted, to extract from the KPZ equation universal, model independent properties one still has to study its long time limit.

To be more precise  than their
size, one must take into account the initial conditions.
Physically  natural options are singled out. One is droplet growth, \textit{e.g.} the Eden cluster, for which the deterministic shape
has a local curvature decaying as $1/t$. Another natural choice would be growth on  a flat substrate, modeled as
that $h(x,0) = 0$. As a consequence $h(x,t)$ will stay flat on average. One also could imagine to evolve for a long 
time span $\tau$ and then 
consider height differences as $h(x,\tau + t) - h(x,\tau)$, which means that the initial height statistics has become stationary and one would have to compare with the KPZ equation for stationary initial conditions.

As will be discussed in more detail in Sect. \ref{sec2}, for the curved case there is an exact solution for the distribution of $h(0,t)$
for all $t> 0$ \cite{AmCoQu11,SaSp11}. From it one deduces that 
\begin{equation}\label{1.8}
h(0,t) \simeq v_\infty t + (\Gamma t)^{1/3} \zeta_{\rm GUE}
\end{equation}
for large $t$, where $v_\infty$ is the asymptotic growth velocity and $\Gamma = \tfrac12 |\lambda| \sigma^4 $.  The random amplitude $\zeta_{\rm GUE}$ has the distribution function
\begin{equation}\label{2.8}
 \mathbb{P}( \zeta_{\rm GUE} \leq s) = \det  (1 - P_s K_{\mathrm{Ai}} P_s)\,.
\end{equation}
The right hand side involves a Fredholm determinant  over
$L^2(\mathbb{R},dx)$, see \cite{ReSi}, Section 17, for Fredholm determinants. $P_s$ is the projection operator onto the interval  $[s,\infty)$ and 
$K_{\mathrm{Ai}}$ the Airy kernel
 \begin{equation}\label{2.9a}
 K_{\mathrm{Ai}}(x,y) = \int_0^\infty du \mathrm{Ai}(x+u)\mathrm{Ai}(y+u)
\end{equation} 
with $\mathrm{Ai}$ the standard Airy function. For the transverse exponent $2/3$, sharp bounds are reported in
\cite{CoHa13}.
For stationary initial conditions an exact expression for the distribution of $h(0,t)$ has been recently accomplished 
\cite{BoCoFeV14} and an asymptotics corresponding to \eqref{1.8} can be deduced.

For the sharp wedge also a replica solution has been achieved \cite{CLDR2010}. By the same method  stationary initial conditions  can be handled \cite{ImSa12,ImSa13}, obtained prior to \cite{BoCoFeV14}.
However for flat initial conditions currently we have to rely on a replica solution only \cite{CLD2011,lDC}. 
 Its structure is much more complicated than in the other two cases.   The replica solution tends to be supported by mathematically rigorous computations on discrete  lattice type models in the KPZ class \cite{OQR14,OQR15}.  What is more certain from such discrete models is  that for large $t$ one should have 
\begin{equation}\label{1.9}
h(t,0) \simeq v_\infty t +  (\Gamma t)^{1/3}\zeta_{\rm GOE}\,,
\end{equation}
where the random amplitude $\zeta_{\rm GOE}$ has the distribution function
\begin{equation}\label{1.10}
 \mathbb{P}( \zeta_{\rm GOE} \leq 2s) = \det  (1 - P_0B_sP_0)
 \end{equation}
with kernel
$B_s(x,y)=\mathrm{Ai}(x+y+s)$, see \cite{Sa06,FeSp06} for this particular representation.

The distribution functions \eqref{2.8}, \eqref{1.10} were obtained first  by Tracy and Widom \cite{TW93} in  apparently completely unrelated problems, namely as the distribution
of the largest eigenvalues of  GUE/GOE random matrices. Just to briefly recall the GUE case, one considers $N\times N$ complex Hermitian
matrices, $A$, with the distribution $Z^{-1} \exp[ - (2N)^{-1}\mathrm{tr}A^2]
dA$. $A$ has the real eigenvalues $\lambda_1 <...<
\lambda_N$, where the factor $N^{-1}$ in front of $\mathrm{tr}A^2$ ensures that their typical distance is of order $1$. For the largest eigenvalue it is proved that 
\begin{equation}\label{2.10}
\lambda_N \simeq 2N + N^{1/3} \zeta_\mathrm{GUE} 
\end{equation}      
 for large $N$. Totally unexpected, the same random amplitude  as for sharp wedge KPZ growth appears, compare with  Eq. (\ref{1.8}).  The GOE case is in complete analogy with real Gaussian symmetric 
matrices.  The GOE/GUE Tracy-Widom  probability distributions can also be represented in terms of special solutions of the Painlev\'e II equation, indicating a link to classical integrable systems.

To return to the numerical/experimental side, if one wants to compare the KPZ predictions \eqref{1.8}, \eqref{1.9} with the data, one first has to carefully determine the value of $\Gamma$. $\lambda$ is measured through the second derivative of the slope-dependent growth velocity. For example, in the case of isotropic growth, one would have $F(\partial_x h) = v_\infty \sqrt{1 + (\partial_x h)^2}$ and hence $\lambda = v_\infty$. The coefficient $\sigma^2$ is more difficult to extract and involves larger uncertainties. In \cite{TS12} it is  used that $\sigma^2=\lim_{x\to\infty}\lim_{t\to \infty} |x|^{-1} {\rm Var}(h(x,t) - h(0,t))$. The sampled distribution for the height at origin at various times plotted in these units should fall on top of each other and agree with the respective Tracy-Widom distributions. \medskip\\
\textbf{1.5 Outline}. 
We first discuss the exact solution with sharp wedge initial data and then
address in more detail the link of lattice type models for interface dynamics to the KPZ equation. 
The second main item is the replica solution. While this method is forced to work with diverging sums, it is still of interest as a tool and in connecting to extensive work on spin glasses and quantum quenches. We will also explain 
a closely related approach which   deals only with convergent sums.  

\section{Exact solution of the KPZ equation for initial sharp wedge}\label{sec2}

On the KPZ level, droplet growth is induced by the sharp wedge initial condition 
\begin{equation}\label{2.1}
h(x,0) = - \delta^{-1}|x| -\log(2\delta)\,,\quad \delta \to 0\,. 
\end{equation}
The solution develops immediately a parabolic shape with superimposed fluctuations, which thus is taken to represent the top part of a droplet. The corresponding initial condition for the stochastic heat equation reads
$Z(x,0) = \delta(x)$.
We note that  
\begin{equation}\label{2.2}
\tilde{Z}(x,t) = \frac{Z(x,t)}{\langle Z(x,t)\rangle} 
\end{equation}
equals the partition function relative to the Brownian bridge
\begin{equation}\label{2.3}
w(s) = b(s) -\frac{s}{t}(b(t) - x)\,, 
\end{equation}
$0 \leq s \leq t$. Expectation w.r.t. the Brownian bridge is denoted by $\mathbb{E}_\mathrm{BB}$. Then
\begin{equation}\label{2.4}
\tilde{Z}(x +a,t) = \mathbb{E}_\mathrm{BB}\Big( \exp\Big[\int_0^t ds \xi(w(s) + a\tfrac{s}{t},s) \Big]\Big)\,. 
\end{equation}
Since $\xi$ is white noise,  $\tilde{Z}(x +a,t)$ has the same distribution as $\tilde{Z}(x,t)$ which implies that $x \mapsto \tilde{Z}(x,t)$
is a stationary process. On the level of the height function this means that $h(x,t) + x^2/2t$ is stationary in $x$.
In fact the exact solution has still a shift by $-\tfrac{1}{24}t$. Thus the object of interest is the deviation, $\eta$,
from the deterministic downwards parabolic profile as
 \begin{equation}\label{2.5}
\eta(x,t) = h(x,t) +\tfrac{1}{2t} x^2 + \tfrac{1}{24} t \,, 
\end{equation}
where $h(x,t)$ is the Cole-Hopf solution of the KPZ equation with sharp wedge initial data. By stationarity it suffices to consider the distribution of $\eta(0,t)$. 

\begin{figure}
\begin{center}\begin{picture}(0,0)\put(185,223){$F_{t}'(s)$}\put(360,7){$s$}\end{picture}\includegraphics[width=125mm]{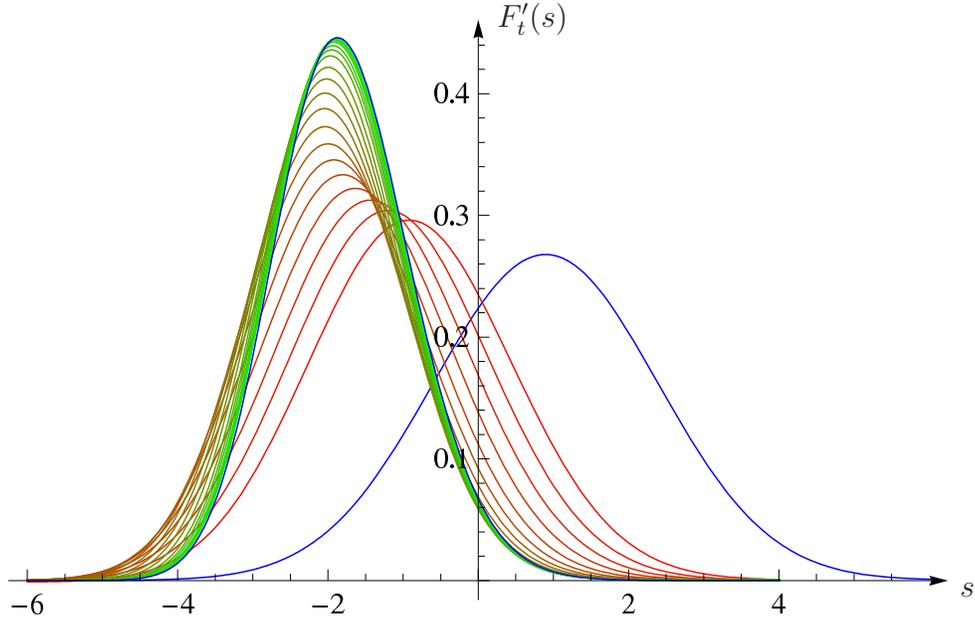}\end{center}
\caption{Probability density function $F_{t}'(s)$ for time $t$ from short times (red, lower curves) to long times (green, upper curves) for  $t$ ranging from 0.25 to 20,000, see \cite{PrSp12} for further details. For $t\to\infty$, the density converges to the GUE Tracy-Widom distribution $F_{\text{GUE}}$ (upper blue curve) and, for $t\to 0$, $F_{t}'(s)$ becomes a Gaussian (rightmost blue curve) with mean and variance increasing respectively as $t^{-1/3}\log t$ and $t^{-1/6}$ on the scale set by the Tracy-Widom distribution.}
\label{fig.2}
\end{figure}

The exact solution was obtained a few years ago \cite{AmCoQu11,SaSp11} and provides an expression for the generating function,
\begin{equation}\label{2.6}
\big\langle \exp{\big[-\exp(\eta(0,t) - \gamma_t s )\big]}  \big\rangle =
\det (1 - P_0K_{t,s}P_0)\,,
\end{equation}
valid for all real $s \in\mathbb{R}$ and $t > 0$, where $$\gamma_t = (t/2)^{1/3}.$$ The right hand side involves a Fredholm determinant  over
$L^2(\mathbb{R},dx)$, 
$P_0$ projects on $\mathbb{R}_+$,  and $K_{t,s}$ is the smoothed Airy kernel  
\begin{equation}\label{2.7}
K_{t,s}(x,y) =  \int_\mathbb{R} du \frac{1}{1 + \mathrm{e}^{- \gamma_t (u -s)}}\mathrm{Ai}(x+u)\mathrm{Ai}(y+u)\,.
\end{equation}
Note that the generating function \eqref{2.6} determines uniquely the distribution function  $F_t(s) = \mathbb{P}(\eta(0,t) \leq s)$.  There is an explicit inversion formula yielding $F_t(s)$,
again involving Fredholm determinants  \cite{AmCoQu11}. Numerically such determinants can be computed very efficiently 
through approximating the integral kernel of $K_{t,s}$ by a finite-dimensional matrix of the form $K_{t,s}(x_i,x_j)|_{i,j = 1,...,n}$ with carefully chosen
base points $\{x_j, j = 1,...,n\}$ \cite{Bo08}. In our application $n = 40$ suffices already to achieve a good precision.
In Fig. \ref{fig.2} we display the probability densities $F_t'(s)$ in dependence on $t$. For short times the required matrix dimension $n$ increases and the numerics becomes slow. In that limit one
returns to the sum (\ref{1.6}) to obtain that the initial fluctuations are Gaussian with a width of order $t^{1/4}$.
At later times a characteristic peak develops which overshoots somewhat to the left and finally settles, on the scale $t^{1/3}$, at the GUE Tracy-Widom distribution. The formula for the limit distribution is deduced from (\ref{2.6}) by substituting 
$\eta(x,t) - \gamma_t s$ by $\gamma_t (\gamma_t^{-1} \eta(x,t) - s)$. Then, taking the limit $t \to \infty$ on both sides, one 
obtains
 \begin{equation}\label{2.9}
\lim_{t \to \infty} \mathbb{P}(\gamma_t^{-1}\eta(0,t) \leq s) = \det  (1 - P_s K_{\mathrm{Ai}} P_s)\,.
\end{equation}

From the exact solution one deduces also some information on finite time corrections. The shape of the distribution
is fairly rapidly attained with the slowest mode being the mean which approaches its limit as $t^{-1/3}$.
Similar behavior is reported from numerical simulations and from the experiment, with a small twist.
As can be seen from Fig. \ref{fig.2}, for the KPZ equation the finite time correction has a negative sign, while in the other cases the sign is positive.
This teaches us that, while the power $t^{-1/3}$ is apparently universal, the prefactor depends on the particular model
\cite{FF11}.

The obvious next step would be to consider the joint distribution of $\eta(0,t),\eta(x,t)$. No closed expression as in \eqref{2.6} is available. 
 But, based on universality, the long time behavior is conjectured with a high level of confidence. 
For simplicity let us consider the joint distribution of $\gamma_t^{-1/3}\eta(0,t),\gamma_t^{-1/3}\eta(x,t)$. We know that these random variables stay
of order 1 as $t\to\infty$. For fixed $x$, they will be perfectly correlated as $t\to\infty$, while they will become 
independent if simultaneously $|x| \to \infty$ sufficiently fast. As anticipated already in the line below \eqref{1.7e},
the interesting scale is 
$x = \mathcal{O}(t^{2/3})$.  \medskip\\
\textbf{Conjecture}. \textit{For every $w \in \mathbb{R}$ it holds that
\begin{equation}\label{2.11}
\lim_{t\to \infty}\mathbb{P}(\gamma_t^{-1/3}\eta(0,t) \leq s_1,\gamma_t^{-1/3}\eta(2\gamma_t^2 w,t)\leq s_2) = \det (1 - K^{(2)}_{s_1,s_2;w})\,,
\end{equation}   
where $K^{(2)}_{s_1,s_2;w}$ is the extended Airy kernel. }\medskip\\
The determinant needs some explanations. $K^{(2)}$ acts on $L^2(\mathbb{R}\times\{1,2\})$
and is given by
\begin{equation}\label{2.12}
K^{(2)}_{s_1,s_2;w} = 
\begin{pmatrix} P_{s_1} & 0 \\
               0 & P_{s_2} \\
\end{pmatrix}\,\begin{pmatrix} K_{\rm Ai} & - \mathrm{e}^{-|w| H}(1- K_{\rm Ai})  \\
              \mathrm{e}^{|w|H} K_{\rm Ai}  & K_{\rm Ai}  \\
\end{pmatrix}\,.
\end{equation} 
$P_s$ is the projection onto $[s,\infty)$ and $H$ is the Airy operator, 
\begin{equation}\label{2.12a}
H = -\frac{d}{dx^2} + x\,.
\end{equation}  
Then $K_{\rm Ai}$ is the projection onto all negative eigenstates of $H$, $K_{\rm Ai} = P(\{H \leq 0\})$,
hence $\mathrm{e}^{-|w| H}(1 - K_{\rm Ai})$ and $ \mathrm{e}^{|w|H} K_{\rm Ai}$ are bounded operators.
More generally one expects that $w \mapsto \gamma_t^{-1/3}\eta(2\gamma_t^2 w,t)$ as a stochastic process converges to the Airy process \cite{PrSp03}, defined through finite dimensional distributions given by expressions generalizing \eqref{2.11} to multi-space points.

With Eq. (\ref{2.11}) one has a tool to study the asymptotic height-height correlation 
\begin{equation}\label{2.13}
\lim_{t \to \infty}\big\langle \big( \gamma_t^{-1/3}\eta(0,t) - \gamma_t^{-1/3}\eta(2\gamma_t^2w,t)\big)^2\big\rangle = g(w)\,.
\end{equation}   
$g(w) \simeq |w|$ for small $ |w|$ and 
\begin{equation}\label{2.14}
g(w) \simeq c^2 - 2 w^{-2}
\end{equation}  
for $|w| \to \infty$ with $c^2$ the variance of the GUE Tracy-Widom distribution, see \cite{TrWi07}. For finer properties one has to rely on the numerical evaluation of the Fredholm determinant in (\ref{2.12}). 

Correlations at the same location but two different times, as \textit{e.g.} the joint distribution of $t^{-1/3}\eta(0,t), t^{-1/3}\eta(0,2t)$,
have been regarded as inaccessible by current methods. But, based on the zero temperature semi-discrete directed polymer, an asymptotic expression has just been posted \cite{Jo15}. Prior it has been argued that under minimal assumptions this joint distribution remains non-degenerate in the limit $t \to \infty$ \cite{Fer11}. 

The proof of \eqref{2.6} is fairly indirect. The general strategy is to start from a discrete growth model, for which sufficient details on the solution with sharp wedge initial data are available. One takes a suitable scaling limit in which the discrete model is known to converge to the KPZ equation. Then the limit of  the generating function for the discrete growth model will yield an identity similar to  \eqref{2.6}. For the sharp wedge one convenient approximating model is the ASEP with step initial conditions \cite{BeGi97}, see Sect. 3.2 for more details. Tracy and Widom  \cite{TW09} derived a Fredholm determinant for the distribution of the $m$-th particle at time $t$, which is sufficiently concise for an asymptotic analysis. A similar strategy has been implemented very recently for stationary initial conditions, $h(x,0) = B(x)$ with $B(x)$ two-sided Brownian motion \cite{BoCoFeV14}. In this case the approximating model is constructed through the theory of Macdonald processes.  

For flat initial conditions, $h(x,0) = 0$, the situation is mixed. For PNG, TASEP, and the zero temperature semi-discrete directed polymer
the large time statistics starting from flat initial conditions have been investigated \cite{Fe04,BFS06,BFPS06,FeSpWe13}. Since these  models do not have an asymmetry parameter, no information on the KPZ equation can be extracted.  
Currently the only candidate is the  asymmetric simple exclusion process \cite{OQR14,OQR15}. However the naturally arising generating functions are not so well suited for an asymptotic analysis.

\section{Approximating the KPZ equation}
\label{sec4}
For models close to a realistic physical description there is always a smallest length scale, say the size of the smallest object attached at, resp. removed from, the interface. Thus the issue arises how such models are related to the KPZ equation. In textbook discussions the focus is on the expansion 
\begin{equation}\label{4.1a}
 \sqrt {1 + (\partial_x h)^2} = 1 + \tfrac{1}{2}(\partial_x h)^2  +\mathcal{O} ((\partial_x h)^4)\,,
\end{equation}
which looks like a small slope expansion. But $h(x)$ has locally the statistics of a Brownian motion and $\partial_xh$ has many spikes.
So perhaps one should first average over some intermediate scale. Through a deeper study of a few model systems, a  perspective different from Taylor expansion has been developed. Namely the KPZ equation is derived from a microscopic model   
in the limit of weak drive. We illustrate with two examples, but first take up the issue of nonlinearities different from quadratic.\medskip\\
\textbf{3.1 Viscous random Hamilton-Jacobi equations}. 
These are stochastic partial differential equations of the form
\begin{equation}\label{SHJ}
\partial_t h =  F(\partial_x h) +\nu\partial_x^2 h +  \xi_\epsilon
\end{equation}
with a general non-linearity $F$.  $\xi_\epsilon$ is space-time white noise smoothed at scale $\epsilon$, see the
paragraph below \eqref{1.7}, and the problem is to make sense of the solution as $\epsilon$ tends to zero. A natural example of such a non-linearity would be the lateral growth
mechanism with $F(\vartheta) = \sqrt{ 1+ \vartheta^2 }$ as in \eqref{4.1a}.
In order to approximate \eqref{SHJ} by the KPZ equation, what  would be done is to expand 
\begin{equation}
F(\vartheta) = F(0) + F'(0) \vartheta + \tfrac12 F''(0) \vartheta^2 + \cdots. 
\end{equation} The $F(0)$ and $F'(0)$ can be
trivially removed by  
height and spatial shifts.   Arguing that the higher than quadratic terms are negligible  one obtains in the new frame the KPZ equation \eqref{1.2} with $\lambda = 
 F''(0)$.

A more convincing approach is to take the ultraviolet cut-off $\e \to 0$ simultaneously  with a weak nonlinearity of size $\sqrt{\e}$. Rescaling as in  \eqref{scal} with $b=1/2$, $z=2$  leads to
\begin{equation}\label{scaling5}
\partial_t h_\e =  \e^{-1} F(\e^{1/2} \partial_x h_\e) + \nu\partial_x^2 h_\e +   \xi_\e.
\end{equation}
Recall  that the
stationary measures of the KPZ equation are of the form 
 \begin{equation}
 h(x)= \sigma B(x) +\vartheta x \,,
 \end{equation}
 where $B(x)$  is a two-sided Brownian motion. The stationary measures
of \eqref{scaling5} are expected to be approximately of the same form, but with $B$ smoothed out on scale $\e$.
One expects that $h_\epsilon(x,t)$ viewed near $x$ is close to a stationary measure with the appropriate slope.
The rapid fluctuations replace $F(\vartheta)$ by  $\bar{F}(\vartheta)$, $\bar{F}$ denoting average with respect to 
such a locally stationary measure. Then, for small $\e$,  
the effective equation reads
\begin{equation}\label{scaling4}
\partial_t \tilde{h}_\e = \e^{-1} \bar{F}(\e^{1/2} \partial_x \tilde{h}_\e) +\nu\partial_x^2 \tilde{h}_\e +
\xi_\e\,.
\end{equation}
$\tilde{h}_\e$ is less wiggly than $h_\e$ and  it is now allowed to take the formal limit to arrive at the KPZ equation \eqref{1.2} with $\lambda = \bar{F}''(0)$ instead of $F''(0)$. Such a claim is proved for polynomial nonlinearities  
in \cite{HQ}.

As an example, one may want to make sense of the quartic KPZ equation 
\begin{equation}\label{SHJ1}
\partial_t h =  (\partial_x h)^4 +\nu\partial_x^2 h +  \xi
\end{equation}
by smoothing out the noise and rescaling  so that all terms are of unit order in the limit. Then  
$\bar{F}(\vartheta) = \langle (\sigma X_\mathrm{G} + \vartheta )^4\rangle$ with $ X_\mathrm{G}$ a standard 
Gaussian and one merely arrives back at the quadratic KPZ equation  with a renormalized $\lambda= \bar{F}''(0) 
=  12\, \sigma^2\langle X_\mathrm{G}^2 \rangle
\neq 0$.  We learn that 
the quadratic KPZ equation is the only case in which it makes sense to have  delta-correlated noise.\medskip\\
\textbf{3.2 Asymmetric simple exclusion process}.
The asymmetric simple exclusion process  models particles on a one-dimensional lattice $\mathbb{Z}$ performing nearest neighbor random walks,
with an asymmetric jump rate and subject to the exclusion rule of at most one particle per site.   
The jump rates are
$p\in[0,1]$ for jumps to the right, \textit{i.e.} from $j$ to $j+1$,  and  $q=1-p$ for jumps to the left,  all jump trials are independent,
and  jumps attempted onto an already occupied 
site are suppressed.  The ASEP height function $h^{\rm ASEP}(j,t)$ is defined by assigning  slope $-1$ to an empty site and slope $1$ to an occupied site, compare with Fig. \ref{fig1}.  $h^{\rm ASEP}(j,t)$ evolves with its own Markovian dynamics, which flips $\vee\leadsto \wedge$ at rate $q$ and $\wedge\leadsto\vee$ at rate $p$. For $q>p$ the general trend is an upwards moving
height function.

In this model there is a remarkable discrete version of the Cole-Hopf transformation \eqref{1.3} as first noted in 
\cite{Ga85}.  
We define the  partition 
function 
\begin{equation}\label{3.9a}
Z^{\rm ASEP}(j,t) = \mathrm{e}^{-(p+q-2\sqrt{pq}) t} \mathrm{e}^{\frac{1}{2}(\log\tau) h^{\rm ASEP}(j,t) }\,,\qquad \tau = p/q\,. 
\end{equation}
It satisfies the stochastic differential-difference equation
\begin{equation}\label{4.2}
dZ^{\rm ASEP} = \sqrt{pq} \Delta Z^{\rm ASEP}dt + Z^{\rm ASEP} dM\,.
\end{equation}
Here $\Delta f (x) = f(x+1) - 2f(x) + f(x-1)$ is the discrete Laplacian and $dM$ is a family of derivative martingales with $\langle dM_j, dM_{j'}\rangle =0$ for $j\neq j'$.  
As in the previous example we adopt the weakly asymmetric limit
\begin{equation}\label{3.9f}
q-p=\e^{1/2}\,.
\end{equation}
 Then $\tfrac12\log\tau \simeq \e^{1/2}=\e^b$ and at diffusive scales, time 
$=\epsilon^{-2} t$, $j =\epsilon^{-1} x 
$,
Eq. \eqref{4.2} converges to the stochastic heat equation  \eqref{1.4}.  Taking logarithms one has established that  the motion of the discrete ASEP height function converges to the Cole-Hopf solution of the KPZ equation \cite{BeGi97}.

Let us specialize from general initial data to step initial conditions, \textit{i.e.} $h^\mathrm{ASEP}(j,0) = - |j|$. Then to leading order $h^\mathrm{ASEP}(0,t) \simeq -\tfrac{1}{4} t$.  For one-sided jumps, at times so short that the discrete nature of $h^\mathrm{ASEP}(0,t)$ is still resolved, numerically one observes a discrete probability measure sitting on top of the Tracy-Widom probability density except for a global shift of order $t^{-1/3}$. The modeling through the KPZ equation
is not so meaningful. Only at sufficiently long times the distributions from both models coincide. On the other hand at sufficiently weak drive there is an intermediate scale on which ASEP and KPZ agree very well. At this point one should worry whether in the limit of weak drive the universal asymptotics might be lost.  A famous example are the long time tails of current correlations in fluids which decay as $t^{-3/2}$ in the physical dimension. On the other hand kinetic theory, a low density approximation, predicts exponential decay. In our context, the exact solution confirms that the KPZ equation just barely maintains the nonlinear effects.\medskip\\
\textbf{3.3 Directed polymer in a random medium}. 
A further instructive example is the directed polymer representation as in \eqref{1.5}.  Probabilistically it is rather obvious how to discretize space-time as $\mathbb{Z}^2$. $b(t)$ is turned into  a discrete time random walk $S_n$ with $n \in \mathbb{N}$,
$S_0 = 0$, and independent increments $S_{n+1} - S_n = \pm 1$ with probability $\tfrac{1}{2}$. White noise is replaced by the family of i.i.d. random variables $\xi(i,j), i,j \in \mathbb{Z}$. Let $S_n$ be an $N$ step random walk. We assign an ``energy" by
\begin{equation}\label{4.3}
E_N(S) = \sum_{n = 1}^N \xi(n,S_n)\,,
\end{equation}
Then the point-to-point partition function reads
\begin{equation}\label{4.4}
Z_\beta^\mathrm{DRP}(N,M) = \sum_{S:(0,0)\mapsto (N,M)} 2^{-N}\exp[\beta E_N(S)]\,.
\end{equation}
We require that $\mathbb{E}\big(e^{\beta \xi(i,j)}\big) < \infty $ at least for small $\beta$. $Z_\beta^\mathrm{DRP}(n,M)$ satisfies a one-step local recursion relation,
\begin{equation}\label{4.5}
Z_\beta^{\rm DRP}(n+1,M) =  \mathrm{e}^{\beta \xi(n+1, M) } \tfrac{1}{2}\big( Z_\beta^{\rm DRP}(n,M+1) + Z_\beta^{\rm DRP}
(n,M-1)\big)\,.
\end{equation}
Taking logarithms, with $h(n,M) = \log Z_\beta^{\rm DRP}(n,M)$,  the recursion becomes
\begin{equation}\label{4.5a}
h(n+1,M) =  \log \big( \tfrac{1}{2}\big(\mathrm{e}^{h(n,M+1)} + \mathrm{e}^{h(n,M-1)}\big)\big) +\beta \xi(n+1, M) \,,
\end{equation}
which
can be  viewed as a particular growth process with additive i.i.d. noise. By analogy one expects,  that
\begin{equation}\label{4.6}
\log Z_\beta^{\rm DRP}(N,0) \simeq v_\infty N + (\Gamma N)^{1/3}\zeta_{\mathrm{GUE}} 
\end{equation}
by KPZ universality. $\lim_{n \to \infty}n^{-1} \log Z_\beta^\mathrm{DRP}(n,0) = v_\infty$  a.s. is a consequence  of subadditivity.
The asymptotic velocity $v_\infty$ can be computed only for few models \cite{Kesten}.  Our understanding of the fluctuations is restricted to the case of $\zeta$ being log-Gamma distributed \cite{seppai}.

To derive the Cole-Hopf representation we choose a lattice spacing $\e$. Brownian motion limit then requires $n = \lfloor
\e^{-2}t\rfloor$,  $\lfloor\cdot\rfloor$ denoting integer part. To have the energy \eqref{4.3} to converge to the continuum energy from \eqref{1.5} one has to scale the noise weak as  $\sqrt{\e}$. This leads to the \textit{intermediate disorder scaling},
\begin{equation}\label{4.7}
\mathrm{e}^{ - \e^{-2}\lambda( (\e/2)^{1/2})  } Z_{(\e/2)^{1/2}} ( \lfloor \e^{-2}t \rfloor,\lfloor \e^{-1}x\rfloor) \,,
\end{equation}
where $\lambda(\beta) = \mathbb{E}\big( e^{\beta \xi}\big)$. This partition function converges to the solution of the stochastic heat equation with $Z(x,0) = \delta(x)$ \cite{AlKhQu14}. The factor $2^{-1/2}$ simply comes from the fact that \eqref{4.5} lives on the even sites of the two dimensional lattice.  It should be emphasized that the result is independent of the distribution of the randomness, only relying on $6$ finite moments of $\min(0,\xi)$.  Again, taking logarithms we see that in these models the
free energy  converges to the respective solution of the  KPZ equation in the intermediate disorder limit.

%
\section{Bethe ansatz and replica solution}
\label{sec3}
As noted already early on  \cite{Ka87}, there are closed evolution equations for the moments of $Z(x,t)$ as defined in 
\eqref{1.5} . Let us consider the
$N$-th moment, now written with $N$ auxiliary Brownian motions, called the replicas. Denoting the white noise average by $\langle \cdot
\rangle$, one arrives at
\begin{equation}\label{3.1}
\big\langle \prod_{j=1}^N Z(x_j,t) \big\rangle =
\big\langle \prod_{j=1}^N\mathbb{E}_{x_j} \big(\mathrm{e}^{\int_0^t ds \xi(b_j(s),s)} Z(b_j(0),0)\big)\big\rangle\,.
\end{equation}
The white noise average can be carried out explicitly and is given by the exponential of 
\begin{equation}\label{3.2}
\frac{1}{2}\int_0^t \int_0^tds_1ds_2 \delta(s_1-s_2)\sum_{i,j=1}^N \delta(b_i(s_1) -b_j(s_2))\,.
\end{equation}
In this form, the It\^{o} discretization amounts to omitting the diagonal term $i = j$. The double time integration reduces trivially to a single one. Hence, using Feynman-Kac backwards, one obtains
\begin{equation}\label{3.3}
\big\langle \prod_{j=1}^N Z(x_j,t) \big\rangle = \langle x_1,...,x_N|\mathrm{e}^{-H_Nt}|Z^{\otimes N}\rangle\,.
\end{equation}
Here $H_N$ is the $N$-particle Lieb-Liniger quantum Hamiltonian on the real line with attractive $\delta$-interaction,
\begin{equation}\label{3.4}
H_N = - \tfrac{1}{2} \sum_{j=1}^N \partial_{x_j}^2 - \tfrac{1}{2} \sum_{i \neq j = 1}^N\delta(x_i - x_j)\,.
\end{equation}
The quantum propagator acts on the product wave function $\prod_{j= 1}^N Z(x_j,0)$, denoted by $Z^{\otimes N}$,
and is evaluated at the point $(x_1,...,x_N)$. Since $Z^{\otimes N}$ is symmetric,
only the restriction of $\exp [-H_N t]$   to the bosonic subspace in $L^2(\mathbb{R}^N)$ is required.
As a result the right hand side of (\ref{3.3}) is a symmetric function, as it should be.

Our representation would work also for the smoothed noise $\xi_\varphi$, the potential $-\delta(x)$ now being replaced by $-\varphi *\varphi(x)$, and also in higher dimensions. But the result is inconsequential unless one has a detailed information on the matrix element in
\eqref{3.3} for all $N$. Fortunately, the attractive $\delta$-Bose gas can be solved through the Bethe ansatz, which would still not be enough. However, for the special case $(x_1,...,x_N)=0$, the formula for the propagator  is sufficiently compact to proceed further.

With this background information we return to the sharp wedge, for which
\begin{equation}\label{3.5}
\langle Z(0,t)^N \rangle = \langle 0|\mathrm{e}^{-H_N t}|0\rangle
\end{equation}
with shorthand $|0\rangle = |0,...,0\rangle$. To compute the propagator we use the eigenfunctions of $H_N$, 
$H_N\Psi_{\underline{\lambda}} = E_{\underline{\lambda}}\Psi_{\underline{\lambda}}$, which are obtained from the Bethe ansatz
\begin{equation}\label{3.6}
\Psi_{\underline{\lambda}}(x_1,...,x_N) = \sum_{p\in\mathcal{P}_N} A_p\prod_{\alpha = 1}^N\mathrm{e}^{
\mathrm{i}\lambda_{p(\alpha)}
x_\alpha}\,,
\end{equation}
where the sum is over all permutations of $1,...,N$. The wave numbers $\lambda_{p(\alpha)}$
turn out to be complex. They are arranged in 
$M$ strings,  $1 \leq M \leq N$, each one containing $n_\alpha$ elements, $\sum_{\alpha =1}^M n_\alpha= N$. A string has momentum $q_\alpha \in \mathbb{R}$ and $n_\alpha$ purely imaginary rapidities as
\begin{equation}\label{3.7}
\lambda^{\alpha,r} = q_\alpha + \mathrm{i}  \tfrac{1}{2}(n_\alpha +1 - 2r)\,,\quad r =  1,...,n_\alpha\,.
\end{equation}
The corresponding energy is
\begin{equation}\label{3.8}
E_{\underline{\lambda}} =  \frac{1}{2} \sum_{\alpha=1}^M n_{\alpha} q_{\alpha}^2
 - \frac{1}{24}\sum_{\alpha=1}^M (n_{\alpha}^3-n_{\alpha})\,.
\end{equation}
We want to compute the generating function
\begin{equation}\label{3.9}
G_t(s) = \big\langle \exp{\big[-\mathrm{e}^{(\eta(0,t) - s )}\big]}  \big\rangle = \sum_{N=0}^\infty \frac{(-1)^N}{N!} 
 \mathrm{e}^{ -N(s -(t/24))}\langle Z(0,t)^N\rangle\,.
\end{equation}
The left average is for a bounded function, hence well-defined.  For the sum, however, the moments should not grow too fast. In actual fact 
\begin{equation}\label{moments}
\langle Z(0,t)^N\rangle \sim \mathrm{e}^{\frac{1}{24} N(N^2-1)t},
\end{equation} and the
sum is ill-defined. Still, the resulting formal computation is instructive and nourishes the hope that one could deal with more general expectations, resp. initial conditions. We follow the steps of \cite{CLDR2010,Dotsenko2010}.

With the appropriate interpretation of the sum over $\underline{\lambda}$, compare with (\ref{3.12}), the Bethe ansatz eigenfunctions are
complete \cite{Oxford}. Hence the $N$-th moment can be written as
\begin{equation}\label{3.10}
\langle Z(0,t)^N \rangle = \sum_{\underline{\lambda}}|\langle 0|\Psi_{\underline{\lambda}}\rangle |^2
\mathrm{e}^{-E_{\underline{\lambda}}t}\,.
\end{equation}
To compute the weights $|\langle 0|\Psi_{\underline{\lambda}}\rangle |^2$ is rather involved \cite{CaCa, Do10}
and can be done systematically only through a proper spectral theory, see \cite{PrSp11}. 
One arrives at
\begin{equation}\label{3.11}
  |\langle 0|\Psi_{\underline{\lambda}}\rangle |^2
  =
  N! \det\left( \frac{1}{\frac12 (n_{\alpha}+n_{\beta})+\mathrm{i}(q_\alpha-q_\beta)} \right)_{\alpha,\beta=1}^M
\end{equation}
and the sum over eigenvalues means
\begin{equation}\label{3.12}
\sum_{\underline{\lambda}} 
= 
\sum_{M=1}^N \frac{1}{M!} \prod_{\alpha=1}^M \sum_{n_\alpha=1}^\infty
\int_\mathbb{R} dq_\alpha (2\pi)^{-1}\delta\big(\sum_{\alpha=1}^M n_\alpha,N\big)\,.
\end{equation}
Hence
\begin{eqnarray}\label{3.13}
 &&\hspace{-30pt}G_{t}(s) = \sum_{M=0}^\infty \frac{(-1)^M}{M!} \prod_{\alpha=1}^M 
 \int_\mathbb{R} dq_\alpha (2\pi)^{-1} \sum_{n_\alpha=1}^\infty (-1)^{n_\alpha-1}
 \mathrm{e}^{n_\alpha^3t/24 - n_\alpha(\frac{1}{2} q_\alpha^2 t  +s)} \notag\\
 &&\hspace{100pt} \times
 \det\left( \frac{1}{\frac{1}{2} (n_{\alpha'}+n_{\beta'})+\mathrm{i}(q_{\alpha'}-q_{\beta'})} \right)_{\alpha',\beta'=1}^M\,,
\end{eqnarray}
with the understanding that the term with $M=0$ equals 1. 

Using 
\begin{equation}\label{3.14}
 \frac{1}{\frac{1}{2} (n_{\alpha}+n_{\beta})+\mathrm{i}(q_\alpha-q_\beta)}
 =
  \int_0^\infty d\omega_\alpha \mathrm{e}^{-( \frac{1}{2} (n_{\alpha}+n_{\beta})+\mathrm{i}(q_\alpha-q_\beta)  )\omega_\alpha} 
\end{equation}
and a simple identity for determinants, one arrives at 
\begin{eqnarray}\label{3.15}
 &&\hspace{-40pt}G_{t}(s) = \sum_{M=0}^\infty \frac{(-1)^M}{M!} \prod_{\alpha=1}^M \int_0^\infty d\omega_\alpha 
 \int_\mathbb{R} dq_\alpha (2\pi)^{-1} \sum_{n_\alpha=1}^\infty (-1)^{n_\alpha-1}
  \notag\\
 &&\hspace{10pt}  \times \mathrm{e}^{n_\alpha^3t/24 - n_\alpha(\mathrm{i} \frac{1}{2} q_\alpha^2 t+ s)}
 \det\left(  \mathrm{e}^{-\frac{1}{2} n_{\alpha'}(\omega_{\alpha'}+\omega_{\beta'}) - \mathrm{i} q_{\alpha'}(\omega_{\alpha'}-\omega_{\beta'}) }  \right)_{\alpha',\beta'=1}^M \,.
\end{eqnarray}
Setting $\gamma_t = (t/2)^{1/3}$, we rescale $s\to \gamma_t s, 
\omega_\alpha \to \gamma_t \omega_\alpha, q_\alpha \to q_\alpha / \gamma_t$
and obtain  
\begin{equation}\label{3.16}
 G_{t}(\gamma_t s) = \det(1-P_0K_{t,s}P_0)\,,
\end{equation}
where the integral kernel of $K_{t,s}$ is given by 
\begin{equation}\label{3.17}
 K_{t,s}(\omega,\omega')
 =
 \int_\mathbb{R} dq (2\pi)^{-1} \mathrm{e}^{\mathrm{i}q(\omega'-\omega)}
 \sum_{n=1}^\infty (-1)^{n-1} \mathrm{e}^{n^3t/24 -\gamma_t n\left( q^2+\frac{1}{2}(\omega+\omega')-s \right)} .
\end{equation}
The projection $P_0$ comes from the restrictions $\omega \geq 0, \omega' \geq 0$. As anticipated, the sum over $n$ is badly divergent. We invoke the identity
\begin{equation}\label{3.18}
 \mathrm{e}^{\lambda^3 n^3/3} = \int_{-\infty}^\infty dy \mathrm{Ai} (y) \mathrm{e}^{\lambda n y} \,.
\end{equation}
The non-uniqueness of the moment problem is reflected by the fact that $\mathrm{Ai}$ in (\ref{3.18}) could be replaced by
some other function out of a huge family without obstructing the identity. Proceeding with this choice,  
we set $\lambda = t^{1/3}/2 = \gamma_t/2^{2/3}$ and shift $y$ to $ y+q^2+\frac{1}{2}(\omega+\omega')$, 
\begin{eqnarray}\label{3.19}
 &&\hspace{-20pt} K_{t,s}(\omega,\omega')
 =
 2^{2/3}\int_{-\infty}^\infty dq(2\pi)^{-1} \mathrm{e}^{\mathrm{i}q(\omega'-\omega)}
  \int dy\mathrm{Ai} \big( 2^{2/3}(y+q^2+\tfrac{1}{2}(\omega+\omega') \big)\nonumber\\  
&&\hspace{120pt}\times \sum_{n=1}^\infty (-1)^{n-1} \mathrm{e}^{-\gamma_t n(s-y)} \,.
\end{eqnarray}
Summing over $n$ and using the identity 
\begin{equation}\label{3.20}
 2^{2/3}  \int_\mathbb{R}dq(2\pi)^{-1} \mathrm{e}^{2\mathrm{i}q u} \mathrm{Ai} \big(2^{2/3}(q^2+x)\big)
 =
 \mathrm{Ai}(x+u) \mathrm{Ai}(x-u) \,,
 \end{equation}
one obtains
\begin{equation}
 K_{t,s}(\omega,\omega')
 =
 \int_\mathbb{R} du \mathrm{Ai}(\omega+u) \mathrm{Ai}(\omega'+u) \frac{\mathrm{e}^{\gamma_t(u-s)}}{1+\mathrm{e}^{\gamma_t(u-s)}} \,,
\end{equation}
in agreement with Eqs. (\ref{2.6}), (\ref{2.7}).

For the joint distribution at two locations, the simplicity of \eqref{3.15} is lost and one can proceed only under further uncontrolled approximations \cite{PrSp11a,PrSp11b, Do13, ImSaSp13}, which do however lead to the expected behavior for $t \to \infty$. For flat initial conditions, instead of setting $(x_1,...,x_N) = 0$ in \eqref{3.3} one has to integrate over all arguments. This corresponds to an initial pure condensate for the $\delta$-Bose gas which is of great interest in the context of quantum quenches \cite{Ca13}.
The analogue of the generating function \eqref{2.6} is derived in \cite{CLD2011}. The reader should be warned that the details are considerably more involved than in the example of this section. For stationary initial conditions the replica solution is accomplished in \cite{ImSa12,ImSa13}.


\section{How to circumvent diverging sums}\label{sec5}

We consider the ASEP with step initial conditions to approximate the KPZ equation with initial sharp wedge and introduce the modified partition function 
\begin{equation}\label{5.1}
\bar{Z}^{\rm ASEP}(j,t) = \mathrm{e}^{\frac{1}{2}(\log\tau) (h^{\rm ASEP}(j,t) +j)} 
\end{equation}
with $\tau = p/q$, $j \in \mathbb{Z}$, and $t \geq 0$. As crucial observation, this family of observables  satisfies 
an analogue of \eqref{3.3},
\begin{equation}\label{5.3}
\big\langle (\tau-1)^{-N}\prod_{\ell=1}^N \big(\bar{Z}^{\rm ASEP}(j_\ell,t) - \bar{Z}^{\rm ASEP}(j_\ell-1,t)\big) \big\rangle = \langle j_1,...,j_N|\mathrm{e}^{-H^{\rm ASEP}_Nt}|Z^{\otimes N}\rangle\,.
\end{equation}
Here $H^{\rm ASEP}_N$ is the lattice $N$ particle operator 
\begin{equation}\label{5.4} 
H^{\rm ASEP}_N= \sum_{\ell=1}^N \big( p\nabla_\ell^- + q\nabla_\ell^+ \big)\,,
\end{equation}
where $\nabla_\ell^\pm f(j_1,...,j_N) =f(j_1,\ldots,j_\ell\pm 1,\ldots,j_N) -f(j_1,\ldots,j_N) $ with boundary conditions
imposed through its domain, which consists of functions $f$
satisfying 
\begin{equation}\label{5.5}
   (p\nabla_{\ell+1}^- + q\nabla_\ell^+ ) f (\vec{j})=0\qquad {\rm whenever } \qquad j_{\ell+1}=j_\ell+1\,.
\end{equation}
In  \eqref{5.3}  it is assumed that $j_1<j_2<\cdots<j_N$. In contrast to the KPZ equation the ``Hamiltonian'' 
$H^{\rm ASEP}_N$ is not a Hermitian operator.
  
As general lore,   such one-dimensional lattice systems, with only nearest neighbor
interactions and conserved number of particles, are solvable through the Bethe ansatz
\cite{TrWi06,BoCoPe15}.  The reality is that most initial conditions lead to unmanageable sums over
the permutation group.  In very special cases one has nice expressions.  A particularly pleasant one
is found in the case of \emph{step} initial data, where initially all the sites $\{1,2,3,\ldots\}$ are 
occupied and sites $\{\ldots,-2,-1,0\}$ are empty.  As established in \cite{BoCoSa13}, the right hand side of \eqref{5.3} can be expressed through an $N$-fold contour integral as
\begin{equation}\label{5.6}
\frac{\tau^{N(N-1)/2} }{(2\pi \mathrm{i})^n}\int_{\mathcal{C}^N}
\prod_{1\leq A<B\leq N} \frac{z_A-z_B}{z_A-\tau z_B}
\prod_{i=1}^{N} \exp\big[ -\frac{z_i(p-q)^2}{(z_i+1)(p+qz_i)}t \big]
\big(\frac{1+z_i}{1+z_i/\tau}\big)^{j_i}  \frac{dz_i}{z_i+\tau}\,,
\end{equation}
where the integration contour $\mathcal{C}$ is a circle around $-\tau$, chosen with small enough radius so that $-1$ is not included, nor is the image of $\mathcal{C}$ under multiplication by $\tau$.  It is important to be 
precise about the contours, because formulas like \eqref{5.6} really code sums of residues.  Such contour integrals have a long history  \cite{HeOp90}.   
Similar formulas appear in  \cite{BoCo12} as solutions to \eqref{3.3}. Note the important difference that unlike \eqref{3.3},
the formula only holds for $j_1<j_2<\cdots<j_N$, and there is some non-trivial work to recover moments from them.

From the solution \eqref{5.6} one has to produce a generating function.  The algebra leads to quantum deformations of the standard exponential function,
\begin{equation}\label{5.7}
e_\tau( z) = {1}/{(z;\tau)_{\infty}} 
\end{equation}
with the Pochhammer symbol
$ (z;\tau)_{\infty}=(1-z)(1-z\tau)(1-z\tau^2)\ldots$. (Most common is the notion of $q$-deformation. It has become a $\tau$-deformation, since $q$ is being used already as jump rate). \cite{BoCoSa13} provide a general route to the moments and the generating 
functions. The result parallels the original derivation of exact formulas in this case by Tracy and Widom \cite{TW09}, and the formulas themselves are similar.
For all complex $\zeta$ except positive reals one has 
\begin{equation}\label{5.8}
\mathbb{E}\Big(e_\tau( \zeta \mathrm{e}^{\frac{1}{2}(\log\tau) (h^{\rm ASEP}(j,t) +j)})\Big) = \det(I+K_{\zeta}^{\rm ASEP})_{L^2(\tilde{\mathcal{C}})}\,,
\end{equation}
where 
\begin{equation}\label{5.9}
K_{\zeta}^{\rm ASEP}(w,w') = \frac{1}{2\pi \mathrm{i}}\int_{\mathcal{D}} \Gamma(-s)\Gamma(1+s)(-\zeta)^s 
\frac{g_{j,t}(w)}{g_{j,t}(\tau^s w)} \frac{(-1)}{\tau^s w-w'}ds
\end{equation}
with
\begin{equation}\label{5.10}
g_{j,t}(w)=
\mathrm{e}^{(q-p)t \tau/(z+\tau)} \left(\tfrac{\tau}{z+\tau}\right)^j\,.  
\end{equation}
The contour $\mathcal{D}$ is a cigar around the positive integers $\{1,\ldots,R\}$ of width $\delta>0$, cutting the 
real axis between $0$ and $1$ and then goes straight up from $R+\mathrm{i} \delta$ to $R+\mathrm{i}\infty$ and likewise straight down from $R-\mathrm{i} \delta$ to $R-\mathrm{i} \infty$,  for  $\delta$  sufficiently small and $R>0$ sufficiently large so that for $w,w'\in \tilde{\mathcal{C}}$, a contour containing $0$ and $-\tau$ but not $-1$, and $s\in \mathcal{D}$, $|q^s w-w'|>0$ and $\left|g_{j,t}(w)/g_{j,t}(q^sw)\right| <B(j,t)$ for some $B(j,t)<\infty$.

One may ask how  the moment problem has been avoided.  Indeed, the moments
$\langle \mathrm{e}^{N h^{\rm ASEP}(j,t)}\rangle$ are expected to grow roughly like
$\mathrm{e}^{C N^2}$, the precise asymptotics being unknown, which is still far too fast to recover the distribution.  Note that in
\eqref{moments} the growth is $\mathrm{e}^{C N^3}$, which demonstrates that the value of exponent for $N$ has little to do with belonging to the KPZ class.  The magical point is however
that because $\log\tau<0$  for left-finite initial data $\mathrm{e}^{\frac{1}{2}(\log\tau) (h^{\rm ASEP}(j,t) +j)} \le 1$.  So it just comes out of this algebra.

To complete the program,
one still has to show that in  the weak drive limit, as discussed in Section 3.2, the generating function \eqref{5.8} converges to the generating function  \eqref{2.6}. This is a difficult piece of asymptotic analysis and still has to be accomplished. In the replica method such a step is avoided because the computation is directly for the KPZ equation with
no approximation scheme.\\\\
\textbf{Acknowledgements}. We are grateful for the generous hospitality at the Institute for Advanced Study, Princeton,
where the first draft was written when both of us participated in the special year on ``Non-equilibrium Dynamics and Random Matrices". We thank Patrik Ferrari for help with the figures.


\end{document}